\documentstyle{article}

\font\tenrm=cmr10 \font\tenit=cmti10 \font\elevenbf=cmbx10
scaled\magstep 1 \font\elevenrm=cmr10 scaled\magstep 1
\newcommand{\be}{\begin{equation}}
\newcommand{\ee}{\end{equation}}

\newcommand{\bea}{\begin{eqnarray}}
\newcommand{\eea}{\end{eqnarray}}

\newcommand{\al}{\alpha}

\newcommand{\hs}{\hspace{5mm}}

\input{epsf}
\begin{document}

\begin{center}
{\elevenbf Universality in a Class of Q-Ball Solutions: An Analytic Approach}

\vglue0.5cm {\tenrm T.A. Ioannidou$^{\dagger }$, A. Kouiroukidis$^{\ddagger }
$, N.D. Vlachos$^{\ddagger }$ \\[0pt]
\vglue0.3cm $^{\dagger }$} {\tenit Mathematics Division, School of
Technology, University of Thessaloniki,Thessaloniki 54124, Greece }\\[0pt]
$^{\ddagger }${\tenit Physics Department, University of Thessaloniki,
Thessaloniki 54124, Greece }\\[0pt]
\vglue0.3cm {\it Emails: ti3@gen.auth.gr\\[0pt]
\hspace{24mm} kouirouki@astro.auth.gr\\[0pt]
\hspace{24mm} vlachos@physics.auth.gr}
\end{center}

{\rightskip=2pc \leftskip=2pc \tenrm\baselineskip=11pt The properties of
Q-balls in the general case of  a sixth order potential have been studied
using analytic methods. In particular, for a given potential, the
initial field value that leads to the soliton solution has been derived and
the corresponding energy and charge have been explicitly evaluated. The
proposed scheme is found to work reasonably well for all allowed values of
the model parameters. \elevenrm\baselineskip=14pt }

\section{Introduction}

A scalar field theory with a spontaneously broken $U(1)$ symmetry may
contain stable nontopological solitons, the so-called Q-balls \cite{col}-%
\cite{Lee}. Q-balls are coherent states of complex scalar fields that carry
a global $U(1)$ charge and can be understood as bound states of scalar
particles which appear as stable classical solutions carrying a rotating
time dependent internal phase. They are characterized by a conserved
nontopological charge $Q$ (Noether charge) that ensures existence and
stability   \cite{drohm}-\cite{belova}.

The concepts associated with these solutions are quite general and occur in
a wide variety of physical contexts \cite{hong}. Q-balls are allowed in
supersymmetric extensions of the standard model that allow flat directions
in the scalar potential. Flat directions in the Minimal Supersymmetric
Standard Model (MSSM) \cite{mssm} have been shown to exist. The conserved
charge is associated with the $U(1)$ symmetries of baryons and leptons,
while, the relevant $U(1)$ fields correspond to either squark or slepton
particles. Thus, Q-balls can be thought of as condensates of a large number
of either squark or slepton particles which can affect baryogenesis via the
Affleck-Dine mechanism \cite{AD} during the post-inflationary period of the
early universe. The Q-ball stability is cosmologically important since if
stable Q-balls are formed in the early universe they can contribute to its
dark matter content. These can be huge balls with charges of order $10^{20}$%
; however, very small Q-balls can also be considered as dark matter
constituents \cite{dark}. Decaying Q-balls can also be of crucial
cosmological significance. If Q-balls decay after the electroweak phase
transition, they can protect baryons from the erasure of baryon number due
to sphaleron transitions. Furthermore, the Q-ball decay may contribute to
dark matter production in the form of the lightest supersymmetric particle,
explaining the baryon to dark matter ratio of the universe \cite{bar}.

Consider the $U(1)$ Goldstone model Lagrangian describing a single complex
scalar field $\phi $ in three spatial dimensions given by 
\begin{equation}
{\cal L}=\frac{1}{2}\partial _{\mu }\phi \,\partial ^{\mu }\bar{\phi}%
-U(|\phi |).  \label{L}
\end{equation}%
$U(|\phi |)$ has a single minimum at $\phi =0$ which is equivalent of
stating that there is a sector of scalar particles (mesons) carrying $U(1)$
charge with mass equal to $\sqrt{\frac{1}{2}U^{\prime \prime }(0)}$. The
corresponding energy functional is given by 
\begin{equation}
E=\int \left( \frac{1}{2}|\dot{\phi}|^{2}+\frac{1}{2}|\nabla \phi
|^{2}+U(|\phi |)\right) d^{3}x  \label{En}
\end{equation}%
while the conserved Noether current  (due to the global $U(1)$ symmetry) is 
\begin{equation}
J_{\mu }=\frac{1}{2i}\left( \bar{\phi}\,\partial _{\mu }\phi -\phi
\,\partial _{\mu }\bar{\phi}\right) 
\end{equation}%
with charge given by 
\begin{equation}
Q=\frac{1}{2i}\int \left( \bar{\phi}\,\partial _{t}\phi -\phi \,\partial _{t}%
\bar{\phi}\right) d^{3}x.  \label{en}
\end{equation}%
The stationary Q-ball solution can be obtained by assuming that 
\begin{equation}
\phi =e^{i\omega t}f(r)  \label{ph}
\end{equation}%
where $f(r)$ is a real radial profile function that satisfies the ordinary
differential equation 
\begin{equation}
\frac{d^{2}f}{dr^{2}}+\frac{2}{r}\frac{df}{dr}=-\omega ^{2}f+U^{\prime }(f)
\label{gene}
\end{equation}%
with boundary conditions $f(\infty )=0$ and $f^{\prime }(0)=0$. In each
case, the effective potential is defined as $U_{eff}(f)=\omega
^{2}f^{2}/2-U(f)$, while, the existence of Q-ball solutions leads to
constraints on the potential $U(f)$ and the frequency $\omega $: (i) The
effective mass of $f$ must be negative, so, by assuming that $U(0)=U^{\prime
}(0)=0$ and $U^{\prime \prime }(0)=\omega _{+}^{2}>0$ one can deduce that $%
\omega <\omega _{+}$. (ii) The minimum of $U(f)/f^{2}$ must be attained at
some positive value of $f$ (say $0<f_{0}<\infty $) and existence of the
solution requires that $\omega >\omega _{-}$ where $\omega
_{-}^{2}=2U(f_{0})/f_{0}^{2}$. Hence, Q-balls exist for all $\omega $ in the
range $\omega _{-}<|\omega |<\omega _{+}$. 

The charge (\ref{en}) and energy (\ref{En}) of a stationary Q-ball solution (%
\ref{ph}) take the simple form 
\begin{eqnarray}
Q &=&4\pi \omega \int r^{2}f^{2}(r)\,dr\   \label{Q} \\
E &=&\frac{1}{2}\omega ^{2}Q^{2}\,+4\pi \int \left( \frac{f^{\prime }{}^{2}}{%
2}+U(f)\right) r^{2}\,dr.  \label{E}
\end{eqnarray}%
Numerical and analytical methods have shown that when the internal frequency
is close to the minimal value $\omega _{-}$ the profile function is almost
constant, implying that the charge (\ref{Q}) is large (thin-wall
approximation). On the other hand, when the internal frequency approaches
the maximal value $\omega _{+}$ the profile function falls off very quickly
(thick-wall approximation). In the thick-wall approximation the behavior of $%
Q$ depends on the particular form of the potential and the number of
dimensions \cite{PC}.

The choice of the potential is not unique since the only requirement is that
the ratio $U(f)/f^{2}$ has a local minimum at some value of $f$ different
from zero. There are several natural types to be considered, two of which
are shown below: 
\begin{eqnarray}
\mbox{I}:\hspace{5mm}U(f) &=&\frac{m_{1}^{2}}{2}f^{2}-\lambda _{1}f^{4}+\mu
_{1}f^{6},  \nonumber \\
\mbox{II}:\hspace{5mm}U(f) &=&m_{3}f^{2}\left( 1-K\log (\lambda
_{3}f^{2})\right) +\mu _{3}f^{2p}.
\end{eqnarray}%
In each case, two of the parameters can be removed by rescaling, thus, the
potentials of type I have one free parameter while potentials of type II
have two for fixed $p$. The type I potential is the simplest allowed one
which is a polynomial in $f^{2}$, while type II mimics the D-flat direction
in MSSM. Here, $p\geq 6$ is some integer that ensures the growth of the
potential for large $f$, but does not destroy the flatness property for
intermediate values of $f$. None of these types are the kind which might be
associated with a renormalizable quantum field theory, but, they are typical
of effective theories incorporating radiative or finite temperature
corrections to a bare potential. In this paper, we shall be dealing with
type I potentials only,  for which $\omega _{+}=m_{1}$ and $\omega _{-}=%
\sqrt{\frac{2m_{1}^{2}\mu _{1}-\lambda _{1}^{2}}{2\mu _{1}}}$ . Stable
Q-balls exist for $\sqrt{\frac{2m_{1}^{2}\mu _{1}-\lambda ^{2}}{2\mu _{1}}}%
<\omega <m_{1}$. The equation of motion (\ref{gene}) is 
\begin{equation}
\frac{d^{2}f}{dr^{2}}+\frac{2}{r}\frac{df}{dr}=\alpha ^{2}f-4\lambda
_{1}\,f^{3}+6\mu _{1}\,f^{5}.  \label{prof}
\end{equation}%
where we have defined $\alpha ^{2}=m_{1}^{2}-\omega ^{2}$.

Q-balls can be studied either analytically  \cite{col,PC,MV}, or numerically 
\cite{drohm}-\cite{AKP}. In a recent work \cite{IKV,IV}, we employed
analytic arguments in order to construct an approximate profile function of
the symmetrized Woods-Saxon type, expected to be valid in the thin-wall
limit. This approximate profile led to an explicit energy-charge relation
which, to our surprise, was found to yield valid results for a region far
exceeding the expected limits of the thin-wall approximation. The
calculation was carried out for a specific form of type I potential i.e. $%
\lambda _{1}=2$, $\mu _{1}=1$ and $m=2$. In this paper, the aforementioned
work is being extended to include the general case of type I potentials
(where the parameters are taken to be arbitrary)  in {\it both} regimes,
thin-wall and thick-wall respectively. This way, the soliton energy and
charge as well as the initial field value $f(0)$  can be accurately derived
as functions of $\alpha $ for the whole allowed range of the model
parameters.

\section{Potential Energy}

From a mechanical point of view, a Q-ball solution describes the motion of a
particle moving with friction in the potential 
\begin{equation}
U_{eff}(f)=-\frac{1}{2}\,\alpha ^{2}\,f^{2}+\lambda _{1}f^{4}-\mu
_{1}f^{6},\hs\hs \lambda _{1},\mu _{1}>0.  \label{p}
\end{equation}%
The corresponding equation of motion is given by (\ref{prof}). Upon
rescaling $r\rightarrow r/\alpha $ and letting $f(r)=f(0)\psi (ar)$ for $%
\psi (0)=1$, equation (\ref{prof}) transforms to 
\begin{equation}
\frac{d^{2}\psi }{dr^{2}}+\frac{2}{r}\frac{d\psi }{dr}=\psi -4\lambda \psi
^{3}+6\mu \psi ^{5}  \label{res}
\end{equation}%
where $\lambda =\frac{\lambda _{1}f(0)^{2}}{\alpha ^{2}}$ and $\mu =\frac{%
\mu _{1}f(0)^{4}}{\alpha ^{2}}$ and the effective potential (\ref{p})
becomes 
\begin{equation}
U(\psi )=-\frac{1}{2}\,\psi ^{2}+\lambda \,\psi ^{4}-\mu \,\psi ^{6}.
\label{pm}
\end{equation}%
After some algebra, it can be shown that equation (\ref{res}) implies that 
\begin{equation}
\int_{0}^{\infty }r^{n-1}\left( \frac{d\psi }{dr}\right) ^{2}dr=\frac{2n}{4-n%
}\int_{0}^{\infty }r^{n-1}\left( -\frac{1}{2}\psi ^{2}+\lambda \psi ^{4}-\mu
\psi ^{6}\right) dr,\hspace{5mm}n\neq 0.
\end{equation}
For the special cases $n=3$ and $n=0$, we get that 
\begin{eqnarray}
\frac{1}{2}\int_{0}^{\infty }r^{2}\left( \frac{d\psi }{dr}\right) ^{2}dr
&=&3\int_{0}^{\infty }r^{2}\left( -\frac{1}{2}\psi ^{2}+\lambda \psi
^{4}-\mu \psi ^{6}\right) dr  \label{der1} \\[3mm]
2\int_{0}^{\infty }r^{-1}\left( \frac{d\psi }{dr}\right) ^{2}dr &=&-\frac{1}{%
2}\psi (0)^{2}+\lambda \psi (0)^{4}-\mu \psi (0)^{6}  \nonumber \\
&\equiv &U(1).  \label{der2}
\end{eqnarray}%
The first condition is Derrick's (or virial) theorem which states that, for
a three dimensional model, the kinetic energy equals three times the
potential energy; the second condition describes the energy dissipation of
the mechanical system.

The mechanical analogue imposes several constraints on $U(1)$. Consider a
particle initially located at the point $\psi (0)=1$ which starts rolling
down the potential wall to eventually stop at the point $\psi (\infty )=0$.
(Here $r$ corresponds to the time variable). Then, the potential energy (\ref%
{pm}) at the origin $U(1)=-\frac{1}{2}+\lambda -\mu $ has to be consumed by
the friction term $\frac{2}{r}\frac{d\psi }{dr}$, therefore $U(1)$ has to be
bounded both from below and above: since for $U(1)$ large the particle will
overshoot the top point while for $U(1)$ small it will not reach it.

The initial potential energy $U(1)$ has to be positive implying that 
\begin{equation}
\mu \leq \lambda -\frac{1}{2}  \label{root}
\end{equation}%
and, also, $U^{\prime }(1)$ has to be positive (attractive \textquotedblleft
force") leading to 
\begin{equation}
\mu \leq \frac{1}{6}\left( 4\lambda -1\right) .  \label{m}
\end{equation}
These conditions are satisfied for $\lambda >1$ when (\ref{m}) holds while
for $1/2<\lambda <1$ when (\ref{root}) holds. Note that for $\lambda <1$  no
Q-ball solutions can be found. Depending on the shape of the potential (i.e.
the actual values of $\lambda $ and $\mu $) the following two distinct cases
occur:

{\it I. Thin-Wall Approximation:} In this case, $\psi (0)$ lies near the
maximum of the effective potential which is deep. Then $U^{\prime
}(1)=-1+4\lambda -6\mu \approx 0$ has to be positive and close to zero (slow
roll) while $U^{\prime \prime }(1)$ is negative (convex region) and of order
unity. The Q-ball solutions lie approximately on the line 
\begin{equation}
\mu =\frac{1}{6}\left( 4\lambda -1\right)  \label{thn}
\end{equation}%
and the initial potential energy depends linearly on $\lambda $: 
\begin{equation}
U(1)=\frac{1}{3}\left( \lambda -1\right) ,\hspace{5mm}\hspace{5mm}\lambda >1.
\label{U1}
\end{equation}

{\it II. Thick-Wall Approximation:} Here, the potential is shallow and the
maxima are high up while $U^{\prime }(1)$ is positive and large (concave
region) and $\mu $ is small. The initial potential energy $U(1)$ increases
with $\lambda $ and reaches its maximal value when $\mu \rightarrow 0$. This
value can be determined numerically and is found to be: $\lambda
_{max}\simeq 4.70137$. The actual functional dependence of $U(1)$ on $%
\lambda $ however, is unknown and needs to be determined.

To that end, let us assume a leading power law behaviour  
\begin{equation}
U(1)=\kappa \lambda ^{n}.  \label{thc}
\end{equation}%
Then, (\ref{pm}) implies that the Q-ball solutions lie on the line 
\begin{equation}
\mu =-\frac{1}{2}+\lambda -\kappa \lambda ^{n}  \label{tel}
\end{equation}%
which determines $\kappa $ in terms of $\lambda _{max}$ (since $\mu (\lambda
_{max})=0$): 
\begin{equation}
\kappa =\frac{2\lambda _{max}-1}{2\lambda _{max}^{n}}.
\end{equation}

\begin{figure}[tbp]
\begin{center}
\epsfxsize=11cm\epsfysize=9cm\epsffile{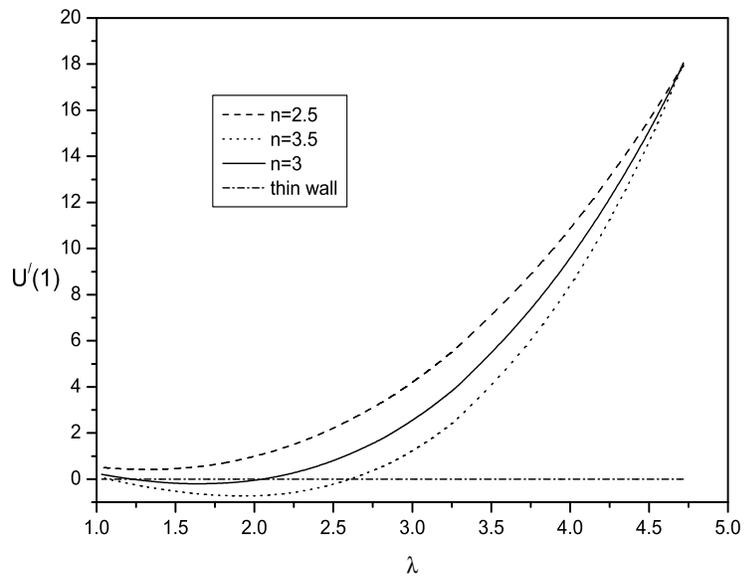}
\end{center}
\caption{The potential energy gradient in terms of $\protect\lambda$  for
different values of $n$.}
\label{vl}
\end{figure}

\begin{figure}[tbp]
\begin{center}
\epsfxsize=11cm\epsfysize=9cm\epsffile{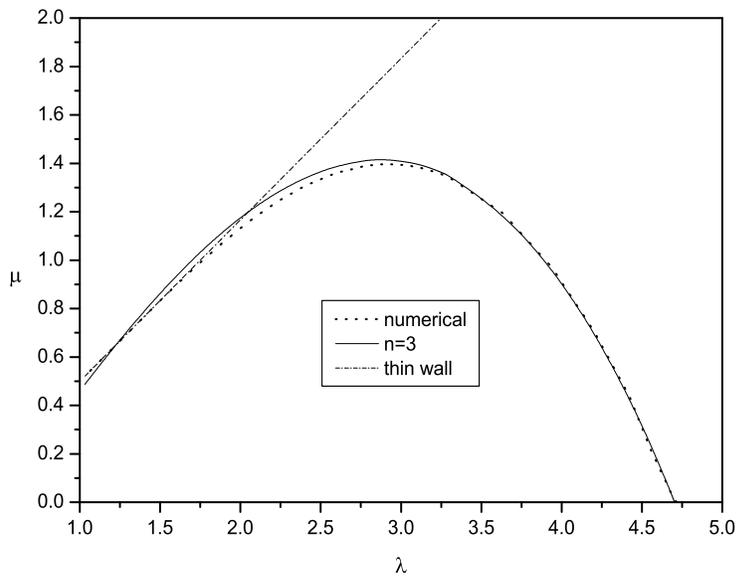}
\end{center}
\caption{The $\protect\mu (\protect\lambda )$ relation for $n=3$ 
against numerical data.}
\label{ml}
\end{figure}

In order to determine the value of $n$ in (\ref{thc}), we require that the
transition from the thick-wall region to the thin-wall one be smooth
implying that, $U^{\prime }(1)$ is small for small values of $\lambda $ ($%
\simeq 1)$, i.e. 
\begin{equation}
U^{\prime }(1)=2(1-\lambda )+6\kappa \lambda ^{n}\approx 0.  \label{t}
\end{equation}%
Its minimum value occurs at the point $\lambda _{\min }=\left( 3n\kappa
\right) ^{\frac{1}{{1-n}}}$. Substituting $\lambda _{\min }$ in (\ref{t})
one gets that $U^{\prime }(1)$ vanishes for $n=2.82187$ in which case $%
\lambda _{\min }=1.54889$ (in agreement to our hypothesis). Assuming that $n$
can only take integer or half integer values (otherwise the structure of the
model on the complex plane would be very complicated), Figure [\ref{vl}]
shows that for $n=3$, the transition is indeed satisfactorily smooth. For
this case, we get that $\kappa =\frac{1}{24.75}$ . This way, we get  the $%
\mu (\lambda )$ relation for the thin-wall (\ref{thn}) and thick-wall (\ref%
{tel}) limits depicted in Figure [\ref{ml}] against numerical data.

\section{Properties of Q-balls}

\begin{figure}[tbp]
\begin{center}
\epsfxsize=11cm\epsfysize=9cm\epsffile{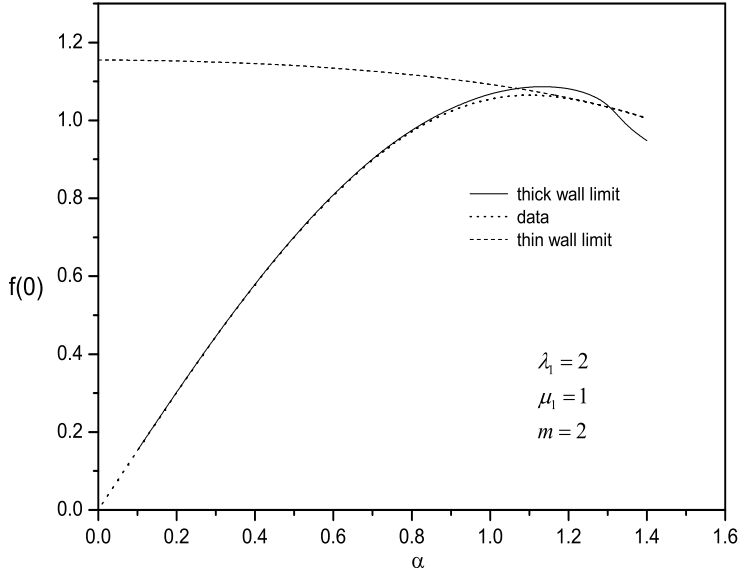}
\end{center}
\caption{Predicted initial field values $f(0)$  against numerical data.
}
\label{f0}
\end{figure}

Using the rescaling formulae of the previous section, the $\alpha $
dependence of the functions $E$, $Q$ and $f(0)$ can now be analytically
determined. The initial field value $f(0)$ that leads to a Q-ball solution
is given by 
\begin{equation}
f(0)=\alpha \sqrt{\frac{\lambda }{\lambda _{1}}},\hspace{5mm}\alpha =\sqrt{%
\frac{\mu }{\mu _{1}}}\frac{\lambda _{1}}{\lambda }.\   \label{f01}
\end{equation}%
Figure [\ref{f0}] depicts predicted values of $f(0)$ in terms of $\alpha $
against numerically calculated ones for sample values $\lambda _{1}=2$, $\mu
_{1}=1$, $m=2$ (as in \cite{IV}).

The corresponding charge (\ref{Q}) and energy (\ref{E}) functionals are
given by: 
\begin{eqnarray}
Q &=&\frac{4\pi \omega }{\alpha }\frac{\lambda }{\lambda _{1}}%
\int_{0}^{\infty }\psi ^{2}\,r^{2}dr  \label{Q1} \\
E &=&\frac{\alpha ^{2}+2\omega ^{2}}{2\omega }\,Q+4\pi \alpha \frac{\lambda 
}{\lambda _{1}}\int_{0}^{\infty }\left( \frac{1}{2}\,\psi ^{\prime
}{}^{2}-\lambda \,\psi ^{4}+\mu \,\psi ^{6}\right) r^{2}dr\ \cdot  \label{E1}
\end{eqnarray}

It was shown in  \cite{IV} that a trial function satisfying the boundary
conditions and having the right asymptotic behavior, is the symmetrized
Woods-Saxon profile 
\begin{equation}
\phi (r)=\frac{c}{\sqrt{1+c_{1}\cosh \left( br\right) }}.  \label{WSsym}
\end{equation}

This function satisfies the exact differential equation 
\begin{equation}
\phi ^{\prime \prime }=\frac{b^{2}}{4}\phi -\frac{b^{2}}{c^{2}}\phi ^{3}+%
\frac{3}{4}\frac{b^{2}}{c^{4}}\left( 1-c_{1}^{2}\right) \phi ^{5}
\end{equation}%
and the approximate one 
\begin{equation}
\phi ^{\prime \prime }+\frac{2}{r}\phi ^{\prime }=\frac{b^{2}}{4}\left( 1-%
\frac{4}{br}\right) \phi -\frac{b^{2}}{c^{2}}\left( 1-\frac{1}{br}\right)
\phi ^{3}+\frac{3b^{2}}{4c^{4}}\left( (1-c_{1}^{2})+\frac{2}{3}\frac{%
c_{1}^{2}}{br}~\right) \phi ^{5}+O\left( \phi ^{7}\right) .
\end{equation}%
Here, $c$, $c_{1}$ and $b$ are arbitrary parameters which need to be
determined in order that (\ref{WSsym}) fits the exact profile function $\psi
(r)$ in the best possible way. Note that only two of the three parameters in
(\ref{WSsym}) are independent since the initial condition $\phi (0)=1$
implies that $c=\sqrt{1+c_{1}}$. 

First note that (\ref{WSsym}) satisfies the following
relations: 
\begin{eqnarray}
\phi ^{\prime }{}^{2} &=&\frac{b^{2}}{4}\,\phi ^{2}-\frac{1}{2}\,\frac{b^{2}%
}{c^{2}}\,\phi ^{4}+\frac{1}{4}\frac{b^{2}}{c^{4}}\left( 1-c_{1}^{2}\right)
\phi ^{6}  \nonumber \\
\phi ^{4} &=&c^{2}\phi ^{2}+c^{2}c_{1}\frac{d}{dc_{1}}\phi ^{2}  \nonumber \\
\phi ^{6} &=&c^{4}\phi ^{2}+2c^{4}c_{1}\frac{d}{dc_{1}}\phi ^{2}+\frac{1}{2}%
c^{4}c_{1}^{2}\frac{d^{2}}{dc_{1}^{2}}\phi ^{2}  \label{pro}
\end{eqnarray}%
in terms of which the charge (\ref{Q1}) and energy (\ref{E1}) functionals
can be explicitly evaluated: 
\begin{eqnarray}
Q &=&\frac{4\pi \omega }{\alpha }\frac{\lambda }{\lambda _{1}}\,\frac{c^{2}}{%
3b^{3}}\,i_{0}  \label{QWS} \\
E &=&4\pi \frac{c^{2}}{3b^{3}}\sqrt{\frac{\mu }{\mu _{1}}}\bigg\{\left( 
\frac{\left( m^{2}+\omega ^{2}\right) }{2\alpha }\frac{\lambda }{\lambda _{1}%
}\sqrt{\frac{\mu _{1}}{\mu }}+\mu c^{4}-\lambda c^{2}-\frac{1}{8}%
b^{2}c_{1}^{2}\right) i_{0}+\left( 2\mu c^{4}-\lambda c^{2}-\frac{1}{4}%
b^{2}c_{1}^{2}\right) c_{1}\frac{di_{0}}{dc_{1}}  \nonumber \\
&&+\frac{1}{2}\left( \mu c^{4}+\frac{1}{8}b^{2}\left( 1-c_{1}^{2}\right)
\right) c_{1}^{2}\frac{d^{2}i_{0}}{dc_{1}^{2}}\bigg\}.  \label{EWS}
\end{eqnarray}%
Here $i_0$ is a function of $c_1$ given by
\begin{equation}
i_{0}\left( c_{1}\right) =\frac{1}{\sqrt{1-c_{1}^{2}}}\,\mbox{arccosh}\left( 
\frac{1}{c_{1}}\right) \left[ \pi ^{2}+\mbox{arccosh}\left( \frac{1}{c_{1}}%
\right) ^{2}\right] ,\hspace{5mm}\mbox{for}\hspace{5mm}c_{1}<1.
\end{equation}
Note that $\lambda _{1}$, $\mu _{1}$, $m$ are fixed external parameters, $%
\lambda $ varies continuously between $1$ and $4.70137$ and $\mu $ is a
known function of $\lambda $ given by (\ref{thn}) or (\ref{tel}) depending
on the regime.

\begin{figure}[tbp]
\begin{center}
\epsfxsize=11cm\epsfysize=9cm\epsffile{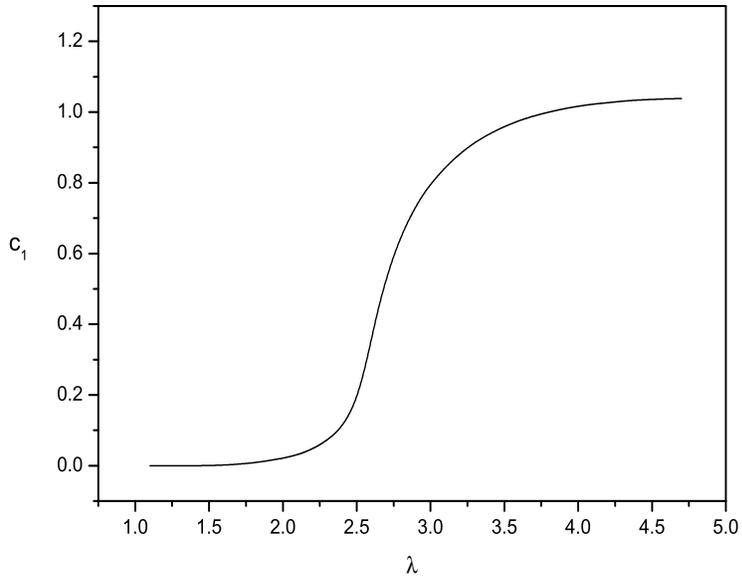}
\end{center}
\caption{The $c_1(\protect\lambda)$ relation obtained from (\protect\ref{c1}%
). }
\label{cl}
\end{figure}

The profile parameters $b$ and $c_{1}$ can  be determined by imposing the
conditions (\ref{der1}) and (\ref{der2}) on $\phi (r)$, that is 
\begin{eqnarray}
\frac{1}{2}\,b^{2}c_{1}^{2}\,c^{2}\int_{0}^{\infty }\frac{dr}{r}\frac{\sinh
^{2}r}{\left( 1+c_{1}\cosh r\right) ^{3}}=U(1)\hspace{5mm}\hspace{5mm}%
\hspace{5mm}\hspace{5mm}\hspace{5mm}\hspace{5mm} &&  \label{fric} \\
\left( \frac{1}{2}-\lambda c^{2}+\mu c^{4}-\frac{b^{2}c_{1}^{2}}{24}\right)
i_{0}+\left( 2\mu c^{4}-\frac{b^{2}c_{1}^{2}}{12}-\lambda c^{2}\right) c_{1}%
\frac{di_{0}}{dc_{1}}+\frac{c_{1}^{2}}{2}\left( \mu c^{4}+\frac{b^{2}}{24}%
\left( 1-c_{1}^{2}\right) \right) \frac{d^{2}i_{0}}{dc_{1}^{2}}=0. &&
\label{duo}
\end{eqnarray}%
Equations (\ref{fric}) can  be written in a compact form 
\begin{equation}
b=\,\frac{1}{c_{1}c}\,\sqrt{\frac{2U(1)}{i_{1}}}  \label{b}
\end{equation}
where 
\begin{equation}
i_{1}=\int_{0}^{\infty }\frac{dr}{r}\,\frac{\sinh ^{2}r}{\left(
1+c_{1}\,\cosh r\right) ^{3}}
\end{equation}
in terms of which (\ref{duo}) becomes
\begin{equation}
\left( \frac{1}{2}-\lambda c^{2}+\mu c^{4}-\frac{1}{12c^{2}}\,\frac{U(1)}{%
i_{1}}\right) i_{0}+\left( 2\mu c^{4}-\frac{1}{6c^{2}}\,\frac{U(1)}{i_{1}}%
-\lambda c^{2}\right) c_{1}\frac{di_{0}}{dc_{1}}+\frac{c_{1}^{2}}{2}\left(
\mu c^{4}+\frac{\left( 1-c_{1}^{2}\right) }{12c_{1}^{2}c^{2}}\frac{U(1)}{%
i_{1}}\right) \frac{d^{2}i_{0}}{dc_{1}^{2}}=0.  \label{c1}
\end{equation}%
Equation (\ref{c1}) determines $c_{1}$ in terms of $\lambda $ and and is
depicted in Figure [\ref{cl}]. Note that the values of $c_{1}$ and $b$
obtained this way are universal since they do not depend on the geometrical
parameters of the potential. Nevertheless, the values of the energy and
charge do depend on the specific form of the potential i.e. the values of $%
\lambda _{1}$, $\mu _{1}$ and $m$. These values tend to infinity in  both
limits since (i)  in the thin-wall limit  $c_{1}\rightarrow 0$  
while (ii)  in the the thick-wall limit $b\rightarrow 0$.

\begin{figure}[tbp]
\begin{center}
\epsfxsize=11cm\epsfysize=9cm\epsffile{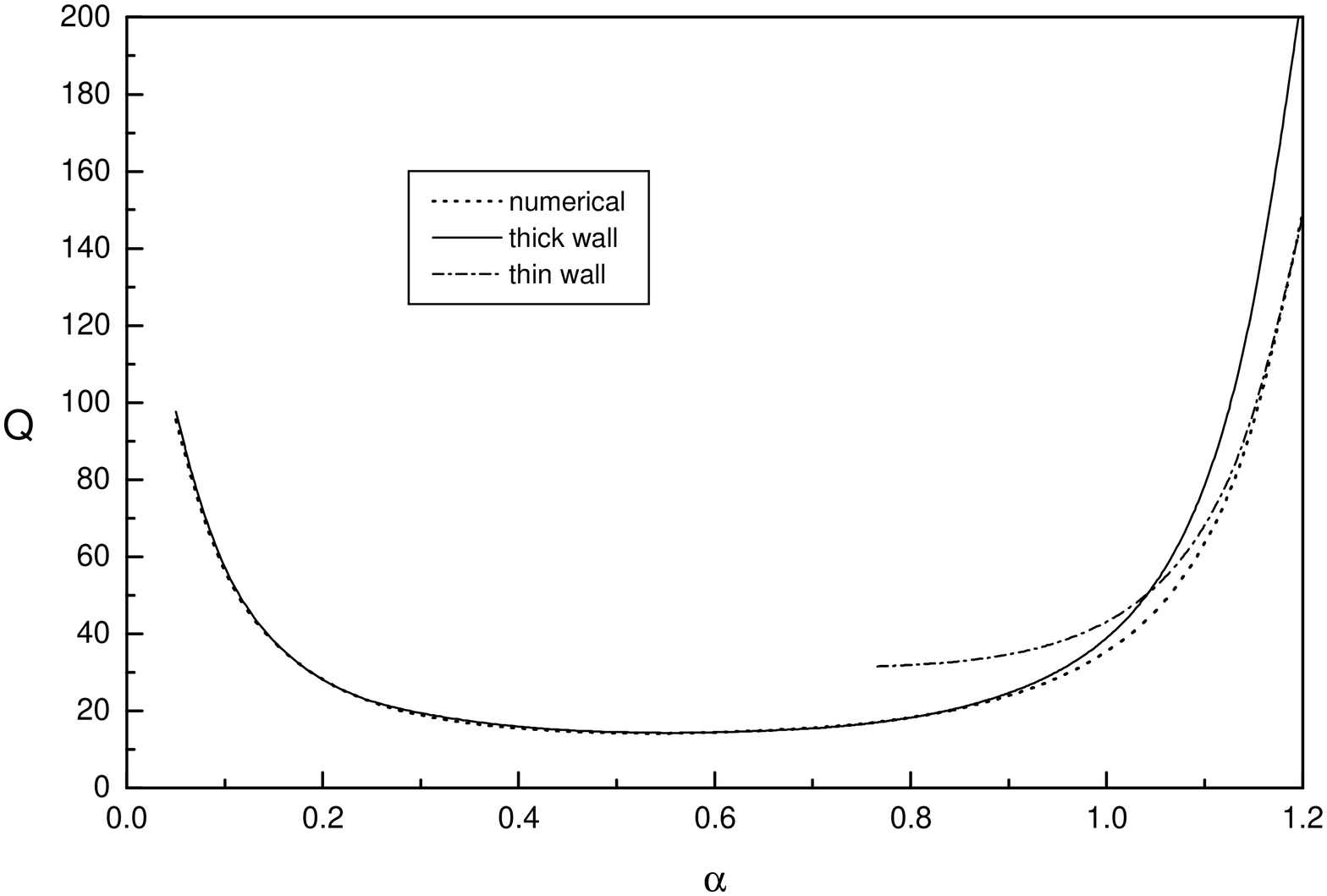 }
\end{center}
\caption{The $\al$ dependence of $Q$ given by (\protect\ref{QWS}%
). }
\label{Qp}
\end{figure}

\begin{figure}[tbp]
\begin{center}
\epsfxsize=11cm\epsfysize=9cm\epsffile{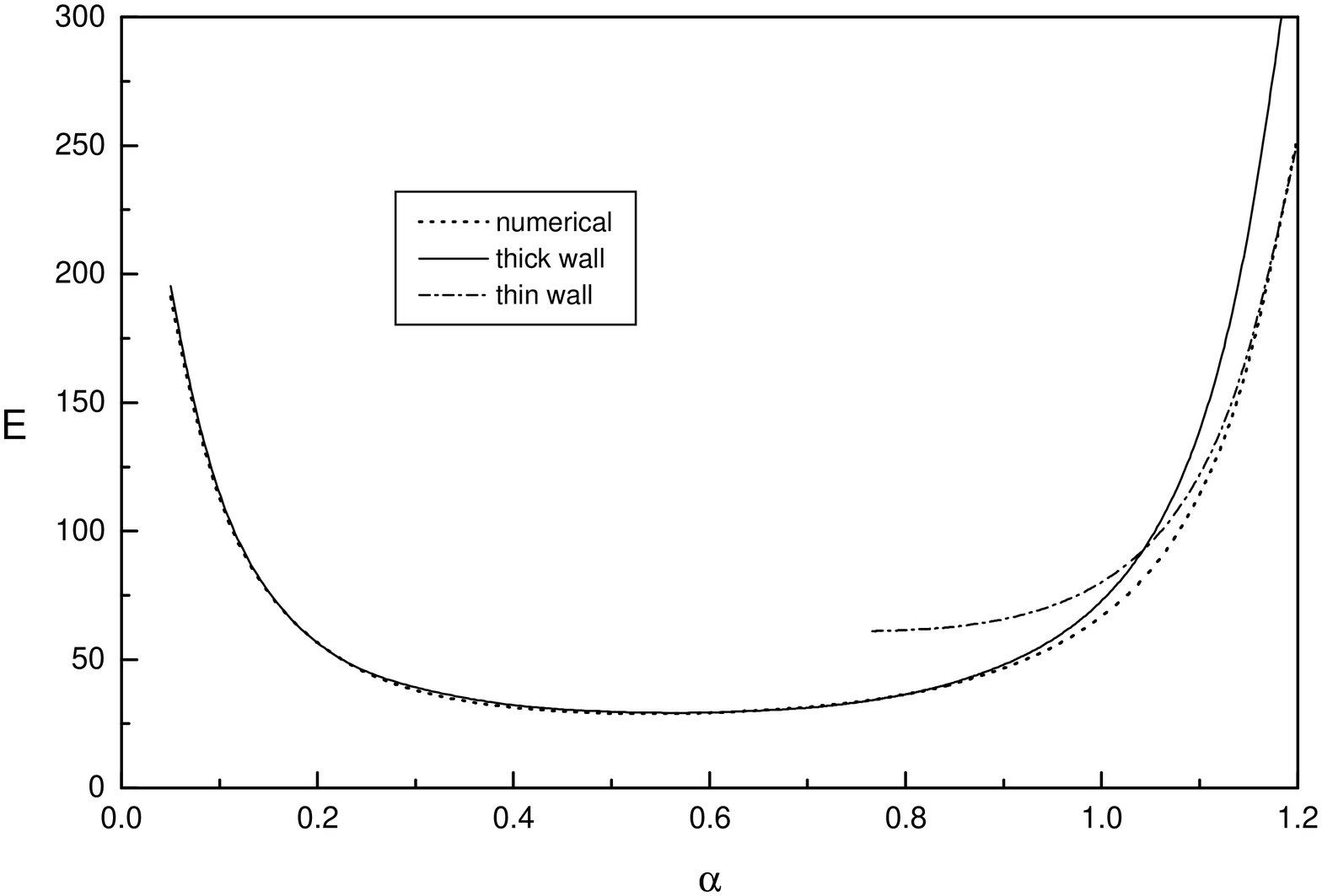}
\end{center}
\caption{The $\al$ dependence of $E$ given by (\protect\ref{EWS}%
). }
\label{Ep}
\end{figure}

Figures [\ref{Qp}] and [\ref{Ep}] presents the $\alpha $ dependence of the
charge (\ref{QWS}) and energy (\ref{EWS}) against values obtained
numerically, for $\lambda _{1}=2$, $\mu _{1}=1$, $m=2$. It is interesting to
realize that the range of validity of the thick-wall approximation is very
wide and gives satisfactory results even in the thin-wall region. It appears
that in this class of theories the Q-balls prefer to be ``thick".

\section{Conclusions}

In this paper we have extended our earlier work \cite{IV} and investigated
phenomenologically relevant properties of Q-balls in a universal way. In
particular, we have addressed the following problem: Given the geometrical
characteristics of the scalar potential find the initial field value that
will lead to a Q-ball solution as well as the corresponding charge and
energy of the soliton. For a particular class of potentials (sixth order
polynomials),  after parameter rescaling, the problem can be tackled in a
universal way. The Q-ball profile can be accurately approximated by
means of a two-parameter symmetrized Woods-Saxon function which can be
analytically calculated in all cases. This scheme is found to yield 
satisfactory results in the whole parameter region so that we do not 
have to rely on approximations like thin-wall or thick-wall. 
The soliton energy and charge 
can subsequently be analytically calculated and compared against numerically
calculated values taken from an earlier work \cite{IKV}.

We believe that a similar line of argument can be applied to study the
profile function and the energy-charge dependence in all types of
potentials. \newline

\end{document}